\documentstyle[preprint,aps,prl]{revtex}

\begin{document}  

%\draft
\thispagestyle{empty}

\null

\vspace{-1cm}
\rightline{\footnotesize hep-th/9806081}
\rightline{\footnotesize NIKHEF 98-016}
\rightline{\footnotesize UvA-WINS-Wisk-98-06}

\vspace{1cm}
\setcounter{footnote}{0}
\begin{center}
{\large{\bf The Matrix Theory S--Matrix}
    }\\[12mm]

\sc{Jan C.\ Plefka$^a$,  Marco Serone$^b$ and
Andrew K.\ Waldron$^a$}\\[5mm]
$^a${\it NIKHEF, P.O. Box 41882, 1009 DB Amsterdam,\\
The Netherlands}\\
{\tt plefka,waldron@nikhef.nl}\\[5mm]
$^b${\it Department of Mathematics, University of Amsterdam,
\\ Plantage Muidergracht 24, 1018 TV Amsterdam,\\
The Netherlands }\\
{\tt serone@wins.uva.nl} \\[18mm]

{\bf Abstract}\\[2mm]
\end{center}
The technology required for
eikonal scattering
amplitude calculations in Matrix theory is developed.
Using the entire
supersymmetric completion
of the $v^4/r^7$ Matrix theory potential
we compute the graviton--graviton scattering 
amplitude and find
agreement with
eleven dimensional supergravity at tree level.

\vfill
\leftline{{\sc June 1998}}

\newpage
\setcounter{page}{1}
 % useful macros
\newcommand {\bea}{\begin{eqnarray}}
\newcommand {\eea}{\end{eqnarray}}
\newcommand {\be}{\begin{equation}}
\newcommand {\ee}{\end{equation}}

\newcommand{\bfa}[1]{{\bf #1}}
\newcommand{\beq}{\begin{equation}}
\newcommand{\eeq}{\end{equation}}

\newcommand\nn\nonumber
\newcommand\eqn[1]{(\ref{#1})}
\newcommand{\NPB}[3]{{Nucl.\ Phys.} {\bf B#1} (#2) #3}
\newcommand{\CMP}[3]{{Commun.\ Math.\ Phys.} {\bf #1} (#2) #3}
\newcommand{\PRD}[3]{{Phys.\ Rev.} {\bf D#1} (#2) #3}
\newcommand{\PLB}[3]{{Phys.\ Lett.} {\bf B#1} (#2) #3}
\newcommand{\JHEP}[3]{{J.\ High Energy Phys.} {\bf #1} (#2) #3}
\newcommand{\ft}[2]{{\textstyle\frac{#1}{#2}}}
\renewcommand{\a}{\alpha}
\renewcommand{\b}{\beta}
\renewcommand{\c}{\gamma}
\renewcommand{\d}{\delta}
\newcommand{\pa}{\partial}
\newcommand{\g}{\gamma}
\newcommand{\e}{\epsilon}
\newcommand{\z}{\zeta}
\newcommand{\m}{\mu}
\newcommand{\n}{\nu}
\newcommand{\x}{\chi}
\newcommand{\p}{\pi}
\newcommand{\s}{\sigma}
\renewcommand{\t}{\tau}
\newcommand{\y}{\upsilon}
\renewcommand{\o}{\omega}
\newcommand{\q}{\theta}
\newcommand{\h}{\eta}
\newcommand{\r}{\rho}
\newcommand{\si}{\sigma}
\newcommand{\ld}[1]{\lambda^\dagger_#1}
\newcommand{\la}[1]{\lambda_#1}
\newcommand{\ps}[2]{(\theta\gamma^{#1 #2}\theta)}
\newcommand{\del}{\partial}
\def\Fss{F\!\!\!\!/\!\!/}
\def\mb{{{\mu}}}
\def\nb{{{\nu}}}
\def\rb{{{\rho}}}
\def\sb{{{\sigma}}}

M-theory, the eleven dimensional quantum theory underlying 
perturbative strings, has 
in recent years headlined dramatic changes
in our understanding of string theory. 
At large distances M-theory reduces (by
definition) to eleven dimensional supergravity.
According to the Matrix theory conjecture of \cite{bfss}
the microscopic degrees of freedom of M-theory are
described by the large $N$ limit of a
{\it quantum mechanical} supersymmetric $U(N)$ Yang--Mills model.
The  model itself
arises, on the one hand, as the regulating theory of the
eleven dimensional supermembrane \cite{dWHN} and on the other as
the short distance description of D0-branes \cite{pol,witt2}.
An essential feature of the model is the existence of
asymptotic particle states carrying the quantum numbers of the eleven
dimensional graviton supermultiplet \cite{bfss,PW}.

A principal test of the Matrix conjecture is the comparison of
scattering amplitudes in the Yang--Mills quantum mechanics
with those of eleven dimensional supergravity.
To date,
typical Matrix theory scattering experiments
involve the comparison of classical gravity
source-probe actions with the
background field effective action of super
Yang--Mills theory in $(1+0)$ dimensions
evaluated on straight line configurations \footnote{See 
\cite{banks} for
an exhaustive list of references.}.
However, a Matrix theory computation yielding true
$S$-matrix elements, depending on momenta and polarizations of
the external particles, has remained elusive.
In this letter we carry out precisely such a computation.

To this end we construct a Matrix theory analogue of the LSZ reduction
formula which relates the $S$-matrix to the background field
expansion of the Matrix theory path integral.
In essence, we have found that the $S$-matrix elements
formed from the 
asymptotic supergraviton states of \cite{PW}
induce exactly the 
boundary conditions in the Matrix path integral satisfied
by straight line diagonal background field configurations.

In fact, in order to obtain the polarization
dependence of scattering amplitudes in Matrix theory,
it is necessary to expand the effective potential
in both bosonic and fermionic background fields. There exist,
scattered in the literature, some partial results for the  
fermionic part of the one loop Matrix theory effective potential
\cite{Kraus97,static}. Here we present the full result which is 
based on the work of \cite{mss1,mss2}. Rather than a Matrix theory 
Feynman diagram {\it tour de force}, all leading D-brane
spin-dependent interactions are obtained 
by a string theory computation employing
the Green-Schwarz boundary state formalism \cite{gregut1}.

Combining the effective potential and our Matrix theory
LSZ reduction formula it is then possible to compute eikonal
$S$-matrix elements. As an example we consider a graviton--graviton 
scattering process and find that the Matrix theory scattering
amplitude agrees with that of eleven dimensional supergravity.

Before presenting our results and formalism, a few remarks are in order.
Throughout this paper we work in the $N=2$ sector of the Matrix
model. Since our computations are, for the time being, restricted to 
the one loop leading terms of Matrix theory which are protected by
supersymmetry, there is no need to take the large $N$ limit.
The demonstrated impressive agreement of supergravity and Matrix theory 
amplitudes at finite $N$ indeed confirms this claim. 
Despite the fact that this agreement is expected by supersymmetry,
our results clearly show that
Matrix theory is aware of the tensorial structure of {\it Lorentz
invariant} eleven dimensional supergravity.
Moreover, the formalism developed in the present 
letter\footnote{A more detailed analysis of the results presented
here will appear in a forthcoming paper.}
permits the computation
of more general scattering amplitudes that will be 
crucial to understand the range of validity of Matrix theory.

\subsection*{LSZ for Matrix Theory}

The $N=2$ Matrix theory Hamiltonian 
\be
H= \ft 1 2 P^0_\mu P^0_\mu + \Bigl ( \ft 12 \vec{P}_\mu \cdot \vec{P}_\mu
+ \, \ft 14 (\vec{X}_\mu \times \vec{X}_\nu)^2
+ \, \ft i 2 \vec{X}_\mu\cdot \vec{\theta}\, \gamma_\mu\times
\vec{\theta}\Bigr )\, 
\label{MTHam}
\ee
is a sum of an
interacting $SU(2)$ part  describing relative motions
and a free $U(1)$ piece pertaining to the
centre of mass. We use a vector notation for the adjoint representation 
of $SU(2)$, $\vec{X}_\mu=(Y^I_\mu,x_\mu)$ and
$\vec{\theta}=(\theta^I,\theta^3)$ (with $I=1,2$ and $\mu=1,\ldots ,9$) 
and may choose a gauge
in which $Y^I_9=0$. 
The model has a potential with
flat directions along a valley floor
in the Cartan sector $x_\mu$ and $\theta^3$. 
The remaining degrees of freedom transverse
to the valley are supersymmetric harmonic oscillators in the variables
$Y^I_\mu$ ($\mu\neq9$) and $\theta^I$.
Upon introducing a large gauge invariant
distance $x=(\vec{X}_9\cdot\vec{X}_9)^{1/2}=x_9$ as the separation of 
a pair of particles, the Hamiltonian \eqn{MTHam} was shown \cite{PW} to
possess asymptotic two particle states of the form 
\be
|p^1_\mu,{\cal H}^1;p^2_\mu,{\cal H}^2\rangle=|0_B,0_F\rangle\,
\ft{1}{x_9}e^{i(p^1-p^2) \cdot
x}e^{i(p_1+p_2)\cdot X^0}
|{\cal H}^1\rangle_{\theta^0+\theta^3}\,|{\cal H}^2
\rangle_{\theta^0-\theta^3}\label{state}
\ee
Here $p^{1,2}_\m$ and ${\cal H}^{1,2}$ are the momenta and
polarizations of the two particles.
The state $|0_B,0_F\rangle$ is the ground state of the superharmonic
oscillators
and the polarization states are the $\underline{44}\oplus\underline{84}
\oplus\underline{128}$ representation of the $\theta^0\pm
\theta^3$
variables, corresponding to the graviton, three-form tensor and gravitino
respectively.

For the computation of scattering amplitudes
one may now form the $S$-matrix in the usual fashion
$
S_{fi}\, =\, \langle {\rm out}| \exp \{-iHT\} |{\rm
in}\rangle \, 
$
with the desired in and outgoing
quantum numbers according to~(\ref{state})
\footnote{The asymptotic states above
are constructed with respect to a large separation 
in the same direction for both in and outgoing
particles, i.e.\ eikonal kinematics.
More general kinematical situations are handled by
introducing a rotation operator into the $S$-matrix \cite{PW1}.}.
The object of interest is then the vacuum to vacuum transition
amplitude
\be
e^{i\Gamma(x'_\mu,x_\mu,\theta^3)}=
{}_{x^\prime_\mu}\langle 0_B,0_F|  \exp \{-iHT\} | 0_B,0_F\rangle_{x_\mu}.
\label{trans}
\ee
Note that the ground states actually depend on the Cartan variables 
$x_\mu$ and $x'_\mu$ through the oscillator mass. Also, both the 
left and right hand sides depend on the operator $\theta^3$.

Our key observation is rather simple. In field theory one is accustomed
to expand around a vanishing vacuum expectation value
when computing the vacuum to vacuum transition amplitude for some
field composed of oscillator modes. In quantum mechanics the idea is
of course exactly the same, and therefore if one is to represent \eqn{trans}
by a path integral one should expand
the super oscillators transverse to the valley about a vanishing vev.
One may then write the
Matrix theory $S$-matrix in terms of a path integral with the stated
boundary conditions
\be
e^{i\Gamma(v_\mu,b_\mu,\theta^3)}=
\int_{{\vec{X}}_\mu=(0,0,x_\mu),\, {\vec{\theta}}=(0,0,\theta^3)}
^{{\vec{X}}_\mu=(0,0,x_\mu'),\, {\vec{\theta}}=(0,0,\theta^3)}
{\cal D}(\vec{X}_\mu,\vec{A},\vec{b},\vec{c},\vec{\theta})\,
\exp(i\,\int_{-T/2}^{T/2}L_{\rm SYM}).
\ee
The Lagrangian $L_{\rm SYM}$ is that of the supersymmetric
Yang--Mills quantum mechanics with appropriate gauge fixing 
to which end we have
introduced ghosts $\vec{b}$, $\vec{c}$ and the Lagrange multiplier
gauge field $\vec{A}$.
The effective action $\Gamma(v_\mu,b_\mu,\theta^3)$
is most easily computed via an expansion about classical trajectories
$X^3_\mu(t)\equiv x_\mu^{\rm cl}(t)
=b_\mu+v_\mu t$ and constant $\theta^3(t)=\theta^3$
which yields the quoted boundary conditions
through the identification $b_\mu=(x'_\mu+x_\mu)/2$ and
$v_\mu=(x'_\mu-x_\mu)/T$.

Up to an overall normalization ${\cal N}$, our LSZ
reduction formula
for Matrix theory is simply
\bea
S_{fi}&=&\delta^9(k'_\mu-k_\mu)e^{-ik_\mu k_\mu T/2}\nonumber\\
&&\hspace{0cm}
\int d^9x' d^9x \,{\cal N}\,
\exp(-iw_\mu x'_\mu +iu_\mu x_\mu)
 \langle {\cal H}^3| \langle {\cal H}^4|e^{i\Gamma(v_\mu,b_\mu,\theta^3)}
 |{\cal H}^1\rangle |{\cal H}^2\rangle
\label{superS}
\eea
The leading factor expresses momentum conservation for the centre of mass
where we have denoted $k_\mu=p_\mu^1+p_\mu^2$  and
$k'_\mu=p_\mu^3+p_\mu^4$ for the in and outgoing particles,
respectively, and similarly for the relative momenta
$u_\mu=(p_\mu^1-p_\mu^2)/2$ and $w_\mu=(p_\mu^4-p_\mu^3)/2$.

In a loopwise expansion of the Matrix theory path integral one finds
$\Gamma(v_\mu,b_\mu,\theta^3)=v_\mu v_\mu T/2+ \Gamma^{(1)}
+\Gamma^{(2)}+\ldots$ of which we consider only the first two terms
in order to compare our results with tree level supergravity.
Inserting this expansion into~(\ref{superS}) and
changing variables $d^9x' d^9x \rightarrow d^9 (Tv) d^9 b$,
the integral over $Tv_\mu$ may be performed via stationary phase.
Dropping the normalization and the overall centre of mass piece  
the $S$-matrix then reads
\be
S_{fi}=e^{-i[(u+w)/2]^2 T/2}
\int d^9b \,
e^{-i q_\mu b_\mu}\,
 \langle {\cal H}^3| \langle {\cal H}^4|
e^{i\Gamma(v_\mu=(u_\mu+w_\mu)/2,b_\mu,\theta^3)}
 |{\cal H}^1\rangle |{\cal H}^2\rangle 
\label{sfi}
\ee
where $q_\mu=w_\mu-u_\mu$. 
It is important to note that in~(\ref{sfi}) the variables
$\theta^3$ are operators
$\{\theta^3_\a,\theta^3_\b\}=\delta_{\a\b}$
whose
expectation between polarization states
$|{\cal H}\rangle$ yields the spin dependence
of the scattering amplitude.

The loopwise expansion of the
effective action should be valid for the eikonal regime, i.e. large
impact parameter $b_\mu$ or small momentum transfer $q_\mu$.
As we shall see below, this limit is dominated by $t$-channel
physics on the supergravity side.

\subsection*{D0 Brane Computation of the Matrix Theory Effective  
Potential}

We must now determine the one-loop 
effective Matrix
potential $\Gamma (v,b,\theta^3)$, namely 
the $v^4/r^7$ term and its supersymmetric completion.
Fortunately the bulk of this computation has already been performed
in string theory by \cite{mss1,mss2} who applied the
Green-Schwarz boundary state
formalism of \cite{gregut1}
to a one-loop annulus computation
for a pair of moving D0-branes. They found
that the leading spin interactions are dictated by a simple zero modes
analysis and their form is, in particular,  scale independent.
This observation allows to extrapolate the results of \cite{mss1,mss2}
to short distances and suggest a Matrix theory description
for tree-level supergravity interactions.

Following \cite{mss1,mss2}, supersymmetric D0-brane interactions
are computed from the correlator
\be
{\cal V}=\frac{1}{16}\int_0^\infty \!\!dt \,
\langle B,\vec{x}=0|e^{-2\pi t\alpha^{\prime}  
p^+(P^--i\partial/\partial x^+)}
e^{(\eta Q^-+\tilde{\eta}\tilde{Q}^-)}e^{V_B}|B,\vec{y}=\vec{b} \rangle
\label{cyl}
\ee
with $Q^-,\tilde{Q}^-$ being the SO($8$) supercharges
broken by the presence of the D-brane, $|B\rangle$
the boundary state associated to D0-branes and 
$V_B=v_i\oint_{\tau=0}\!d\sigma\left(X^{[1}\partial_{\sigma}X^{i]}
+\frac{1}{2}S\,\gamma^{1i}S\right)$
is the boost operator where the direction 1 has to be identified
with the time (see \cite{mss1,mss2} for details). Expanding 
(\ref{cyl}) and using the results in
section four of \cite{mss2}, one 
finds the following compact form for the
leading one-loop Matrix theory potential (normalizing to one the 
$v^4$ term and setting $\alpha^\prime=1$)
\bea
{\cal V}_{\rm 1-loop}&=&
\Bigl [ v^4 + 2i\, v^2\,v_m\ps{m}{n}\, \del_n
-2\, v_p v_q
\ps{p}{m}\ps{q}{n}\,\del_m \del_n\nonumber\\
&&\quad -\frac{4i}{9}\, v_q\ps{q}{m}\ps{n}{k}\ps{p}{k}\,\del_m\del_n\del_p
\nonumber \\
&&\quad + \frac{2}{63}\, \ps{m}{l}\ps{n}{l}\ps{p}{k}\ps{q}{k}\,
\del_m\del_n\del_p \del_q\Bigr ]\, \frac{1}{r^7} \label{pot}
\eea
where $\theta=(\eta^a, \tilde \eta^{\dot a})$ should be identified with 
$\theta^3/2$ of the last section. The general structure of this potential
was noted in \cite{harv} and its first, second and last 
terms were calculated
in \cite{dkps},\cite{Kraus97} and \cite{static} respectively.
Naturally it would be interesting to establish the supersymmetry
transformations of this potential; for a related discussion see
\cite{pss1}.

\subsection*{Results}

Our Matrix computation is completed by taking the 
quantum mechanical expectation of
the effective potential 
\eqn{pot} between the polarization states of
\eqn{sfi}. Clearly one can now study any amplitude involving
gravitons, three--form tensors and gravitini.
We choose to compute a $h_1 + h_2 \rightarrow h_4 + h_3$
graviton-graviton process, and thus
prepare states
\bea
|{\rm in}\rangle&=& \ft{1}{256}\, h^1_{mn}\,
(\ld{1}\gamma_{m}\ld{1})(\ld{1}\gamma_n\ld{1})
\, h^2_{pq}\, (\ld{2}\gamma_{p}\ld{2})(\ld{2}\gamma_q\ld{2})\,
|-\rangle \, .
\label{in}
\nn\\
\langle{\rm out}|&=& \ft{1}{256}\, \langle -|\,
h^4_{mn}\, (\la{1}\gamma_{m}\la{1})(\la{1}\gamma_n\la{1})
\, h^3_{pq}\, (\la{2}\gamma_{p}\la{2})(\la{2}\gamma_q\la{2})
\label{out}
\eea
Note that (following \cite{PW}) we have complexified the 
Majorana centre of mass and Cartan spinors $\theta^0$ and $\theta^3$ 
in terms of 
$SO(7)$ spinors $\lambda^{1,2}=(\theta^0_+\pm\theta^3_+
+i\theta^0_-\pm i\theta^3_-)/2$ 
where $\pm$ denotes projection with respect to $\gamma_9$.
Actually the polarizations in \eqn{out}
are seven dimensional but may be generalized to the nine
dimensional case at the end of the calculation.
We stress that these manoeuvres are purely technical and our 
final results are $SO(9)$ covariant.  
The creation and destruction operators $\lambda^\dagger_{1,2}$ and 
$\lambda_{1,2}$ annihilate the states
$\langle-|$ and $|-\rangle$, respectively.

The resulting one loop
eikonal Matrix theory graviton-graviton scattering amplitude 
is comprised of 68 terms and (denoting e.g.\ 
 $(q h_1h_4 v)=q_\mu h^1_{\mu\nu}
h^4_{\nu\rho}v_\rho$ and $(h_1 h_4)=h^1_{\mu\nu}h^4_{\nu\mu}$)
is given by

\begin{eqnarray}
{\cal A}&\,\,= \,\,\frac{\textstyle 1}{\textstyle q^2}\,\,\Biggl\{\,\,& 
\ft12(h_1 h_4)(h_2 h_3) v^4 
+ 2\Bigr[(q h_3 h_2 v) (h_1 h_4) 
      - (q h_2 h_3 v) (h_1 h_4)\Bigr] v^2 
\nn\\&&\hspace{-.23cm}
+  (vh_2v) (qh_3q)(h_1 h_4) 
+  (vh_3v) (qh_2q)(h_1 h_4) 
- 2(qh_2v) (qh_3v)(h_1 h_4) 
\nn\\&&\hspace{-.23cm}
- 2 (qh_1h_4v) (qh_3h_2v)
+ (qh_1h_4v) (qh_2h_3v) 
+ (qh_4h_1v) (qh_3h_2v)  
\nn\\&&\hspace{-.23cm}
+ \ft{1}{2}\Bigl [(qh_1h_4h_3h_2q)
-  2(qh_1h_4h_2h_3q) 
+  (qh_4h_1h_2h_3q) 
-  2(qh_2h_3q)(h_1 h_4) \Bigr ] v^2 
\nn\\&&\hspace{-.23cm}
-  (qh_2v)  (qh_3q)  (h_1h_4)
+  (qh_2q)  (qh_3v)  (h_1h_4)
-  (qh_1q)  (qh_2h_3h_4v) 
+  (qh_1q)  (qh_3h_2h_4v) 
\nn\\&&\hspace{-.23cm}
-  (qh_4q)  (qh_2h_3h_1v)
+  (qh_4q)  (qh_3h_2h_1v)
-  (qh_1v)  (qh_4h_2h_3q) 
+  (qh_1v)  (qh_4h_3h_2q) 
\nn\\&&\hspace{-.23cm}
-  (qh_4v)  (qh_1h_2h_3q)
+  (qh_4v)  (qh_1h_3h_2q)
+  (qh_1h_4q)  (qh_2h_3v) 
-  (qh_1h_4q)  (qh_3h_2v)          
\nn\\&&\hspace{-.23cm}
+\ft18 \Bigl[ 
   (qh_1q)  (qh_2q)  (h_3h_4) 
+2  (qh_1q)  (qh_4q)  (h_2h_3)
+2  (qh_1q)  (qh_3q)  (h_2h_4) 
\nn\\&&\hspace{-.23cm}
+  (qh_3q)  (qh_4q)  (h_1h_2) \Bigr]
+ \ft12\Bigl[
     (qh_1q)  (qh_4h_2h_3q) 
-    (qh_1q)  (qh_2h_4h_3q) 
\nn\\&&\hspace{-.23cm}
-    (qh_1q)  (qh_4h_3h_2q) 
-    (qh_4q)  (qh_1h_2h_3q)
+    (qh_4q)  (qh_1h_3h_2q)
-    (qh_4q)  (qh_2h_1h_3q) \Bigr]
\nn\\&&\hspace{-.23cm}
+ \ft14\Bigl[
   (qh_1h_3q)  (qh_4h_2q) 
+  (qh_1h_2q)  (qh_4h_3q) 
+  (qh_1h_4q)  (qh_2h_3q) \Bigr]
\, \Biggr\}
\nn\\&&\hspace{-.73cm}
+\,\, \Bigl[h_1 \longleftrightarrow h_2\, , \, h_3 \longleftrightarrow h_4
\Bigr]
\label{Ulle}
\end{eqnarray}

We have neglected all terms within the curly brackets proportional
to $q^2\equiv q_\mu q_\mu$, i.e. those that cancel the $1/q^2$ pole.
These correspond to contact interactions in the D0 brane 
computation, whereas this calculation is
valid only for non-coincident branes.

\subsection*{$D=11$ Supergravity}

The above leading order result for eikonal scattering in Matrix theory
is easily shown to agree with the corresponding eleven
dimensional field theoretical amplitude.
Tree level graviton--graviton scattering is dimension independent
and has been computed in \cite{San}.
We have double checked that work by a type IIA string theory computation
and will not display the explicit result here
which depends on eleven momenta $p^i_M$ (with $i=1,\ldots,4$)
and polarizations
$h^i_{MN}$ subject to the de Donder gauge condition
$p^i_N h^i{}_M{}^N-(1/2)p^i_M h^i{}_N{}^N=0$ (no sum on $i$).
Matrix theory, on the other hand, is
formulated in terms of on shell degrees of freedom only, namely
transverse physical polarizations and euclidean nine-momenta.

Going to light-cone variables for the eleven momenta $p^i_M$ 
we take the case of vanishing $p^-$ momentum exchange
\footnote{We
denote $p_\pm=p^\mp
=(p^{10}\pm p^0)/\sqrt{2}$ and our metric convention is
$\eta_{MN}={\rm diag}
(-,+\ldots,+)$.}, i.e. the scenario of our Matrix computation,
\bea
p_M^1=(-\ft12\,(v_\mu-q_\mu/2)^2 ,\, 1\, ,
v_\mu-q_\mu/2 )
&\quad& p_M^2=(-\ft12\, (v_\mu-q_\mu/2)^2 ,\, 1\, , 
-v_\mu+q_\mu/2) \nn\\
p_M^4=(-\ft12\, (v_\mu+q_\mu/2)^2 ,\, 1\, ,
v_\mu+q_\mu/2) &\quad & p_M^3=(-\ft12\,
(v_\mu+q_\mu/2)^2 ,\, 1\, ,
-v_\mu-q_\mu/2) \, .\label{kinematics}
\eea
By transverse Galilean invariance we have set to zero the nine 
dimensional centre of mass
momentum. We measure momenta in units of $p_-$ which we set to one.
For this kinematical situation conservation of $p_+$ momentum
clearly implies $v_\mu q_\mu=0$. 
Note that the vectors $u_\mu$ and $w_\mu$
of~(\ref{superS}) are simply
$u_\mu=v_\mu-q_\mu/2$ and $w_\mu=v_\mu+q_\mu/2$

We reduce to physical
polarizations by using the residual gauge freedom to set $h^i_{+M}=0$
and solve the de Donder gauge condition
in terms of the transverse traceless
polarizations $h^i_{\m\n}$ for which
one finds $h^i_{-M}=-p^i_\n h^i_{\n M}$.

Agreement with the Matrix result \eqn{Ulle} is then achieved by
taking the eikonal limit $v_\mu>>q_\mu$ of the gravity amplitude
in which the $t$-pole contributions
dominate\footnote{In the above parametrization,
the Mandelstam variables are $t=q_\mu^2=-2p^1_M p_4^M$,
$s=4v_\mu^2+q^2_\mu=2p^1_M p_2^M$ and $u=4v_\mu^2=-2p^1_M p_2^M=s-t$.}.
One then reproduces exactly \eqn{Ulle} as long as any pieces cancelling the
$t$-pole (i.e. the aforementioned $q^2$ terms) are neglected.

Although we have only presented here a Matrix scattering amplitude
restricted to the eikonal regime, we nevertheless believe 
the agreement found is rather impressive. 

\subsection*{Acknowledgements}

We thank B. de Wit, S. Moch, K. Peeters and J. Vermaseren
for discussions. Our computation made extensive use of the 
computer algebra system FORM \cite{Jos}.

\end{document}